\documentclass[prl,twocolumn,showpacs,twoside,showpacs]{revtex4}

\usepackage{graphicx}
\usepackage{dcolumn}
\usepackage{bm}
\usepackage{color}

\begin{document}

\bibliographystyle{prsty}
\title{Resolving neutrino mass hierarchy from
supernova (anti)neutrino-nucleus reactions}
\author{D. Vale$^1$}
\author{N. Paar$^1$}
\email{npaar@phy.hr}
\affiliation{$^1$Physics Department, Faculty of Science, University of Zagreb, 
Croatia}

\date{\today}
\begin{abstract}
We introduce a hybrid method to determine neutrino mass hierarchy by simultaneous 
measurements of detector responses induced by antineutrino 
and neutrino fluxes from accretion and cooling phase of type II supernova.
The (anti)neutrino-nucleus cross sections for $^{12}$C, $^{16}$O, $^{56}$Fe and $^{208}$Pb
are calculated in the framework of relativistic nuclear energy density functional and weak Hamiltonian,
while the cross sections for inelastic scattering on free protons, $p(\bar{\nu}_{e},e^{+})n$, are obtained 
using heavy-baryon chiral perturbation theory. 
The simulations of (anti)neutrino fluxes emitted from a protoneutron star in a core-collapse supernova 
include collective and Mickheev-Smirnov-Wolfenstein effects inside star.
The emission rates of elementary decay modes of daughter nuclei are calculated for normal and
inverted neutrino mass hierarchy. It is shown that simultaneous use of (anti)neutrino detectors
with different target material and time dependence of the signal
allow to determine the neutrino mass hierarchy from the ratios 
of $\nu_e / $ $\bar{\nu}_e$ induced particle emissions. The hybrid method favors
detectors with heavier target nuclei ($^{208}$Pb) for the neutrino sector,
while for antineutrinos the use of free protons and light nuclei
($\text{H}_2\text{O}$ or $\text{-CH}_2\text{-}$) represent appropriate choice.
\end{abstract}
\maketitle
\date{today}

Over the past years a considerable
progress has been achieved in constraining the mixing parameters in
neutrino oscillation framework \cite{Mak.62, Pon.67}, based on various experiments on 
atmospheric, solar, and terrestrial neutrinos \cite{Gon.08}. 
It is now well established that neutrinos have nonvanishing rest masses and that the flavor states
$\nu_e$, $\nu_{\mu}$, and $\nu_{\tau}$ are quantum mechanical mixtures of the 
vacuum mass eigenstates $\nu_1$, $\nu_2$, and $\nu_3$~ \cite{Stu.10}.
However, currently existing data do not determine neutrino mass hierarchy, i.e.,
the sign of mass squared difference $\Delta m_{31}^2=m_3^2-m_{1}^2$. 
In the case of $\Delta m_{31}^2 >$ 0 one refers to normal mass hierarchy
(NMH), while $\Delta m_{31}^2 <$ 0 corresponds to inverted mass 
hierarchy (IMH). Although a number of techniques has been proposed to
resolve the neutrino mass hierarchy, to date this question still remains open 
{and represents an important challenge in physics.}
Recent approaches to resolve the neutrino mass 
hierarchy include methods based on reactor neutrinos \cite{Pet.02,Li.13,Cap.14}
different baseline experiments \cite{Ish.05}, Earth matter effects on supernova
neutrino signal~\cite{Lun.03,DDM.08}, spectral swapping of supernova neutrino
flavors~\cite{Dua.07}, rise time of supernova $\bar{\nu}_e$ light 
curve \cite{Ser.12}, analysis of meteoritic supernova material \cite{Mat.12},
and detection of atmospheric neutrinos in sea water or ice \cite{Win.13}.
 
In this Letter we introduce {a hybrid method} to determine the neutrino 
mass hierarchy, based on type II supernova neutrino and antineutrino 
reactions with atomic nuclei, including, $^{12}$C, $^{16}$O, $^{56}$Fe, $^{208}$Pb, 
and free protons.
The aim is to explore how $\nu_e$ and $\bar{\nu}_e$ detectors, based on various
nuclei as target material, can provide a source of information that
is needed to determine the neutrino mass hierarchy.
{Since in the case of supernova event SN1987A mainly the $\bar{\nu}_e$ sector 
of the response has been detected, the role of neutrinos and their relevance for
understanding their underlying fundamental properties remain vastly unknown.}
While most of supernova detectors {based on nucleon or nuclear targets} 
are primarily sensitive to antineutrinos,
Helium and Lead Observatory (HALO) that was recently developed is sensitive
to neutrinos  through charged current (CC) interaction mainly 
with $^{208}$Pb \cite{Dub.08}.
{For the purpose of the present study, microscopic theory framework 
based on relativistic nuclear energy density functional and weak 
Hamiltionian is employed in description of nuclear properties,
neutrino induced excitations, and weak interaction transition matrix elements \cite{Paa.11,Paa.13}. 
In order to account for the $\nu_e$($\bar{\nu}_e$) - induced events in detector, primary
particle decay modes of daughter nuclei are described. In addition,
to determine responses in water and mineral oil,
$\bar{\nu}_e$- free proton cross sections are calculated in the framework introduced in
Ref. \cite{Rah.12}, based on chiral perturbation theory. }

Neutrino and antineutrino fluxes from the core-collapse supernova provide a
powerful tool to probe various neutrino properties, as well as the dynamics 
of star explosion \cite{Sam.93, Das.10}. They represent a kind of a fingerprint for
the events occurring in stellar collapse and can be used for better understanding 
of neutrino flavor transformations in regions of high neutrino densities due to 
neutrino-neutrino interaction \cite{Sam.93, Dua.06, Das.08, Das.10, Mir.11, Vaa.11}, 
and {transitions occurring in} matter 
resonance layers of star due to Mickheev-Smirnov-Wolfenstein (MSW)
effect \cite{Wol.79, Dig.00, Smi.03}. Region 
dominated by collective effects and two MSW resonant 
regions are {spatially well separated, so probabilities} 
of flavor transition can be calculated separately and multiplied. {Both effects depend on} 
neutrino mass hierarchy and change the shape of initial $\nu_e$($\bar{\nu}_e$) 
spectra. Spectral splits and swaps in the $\nu_e$($\bar{\nu}_e$) spectra caused by 
collective effects {also depend on the flux} ordering among different neutrino species, 
which are generally different in accretion and cooling supernova phase. 
{Arrival of shock waves} in the outer layers of a star can leave a mark on $\nu_e$($\bar{\nu}_e$) spectra, 
even cause a non-adiabatic conversion, and multiple MSW effects \cite{Vaa.11}. 

\begin{figure}
\centerline{
\includegraphics[scale=0.35,angle=0]{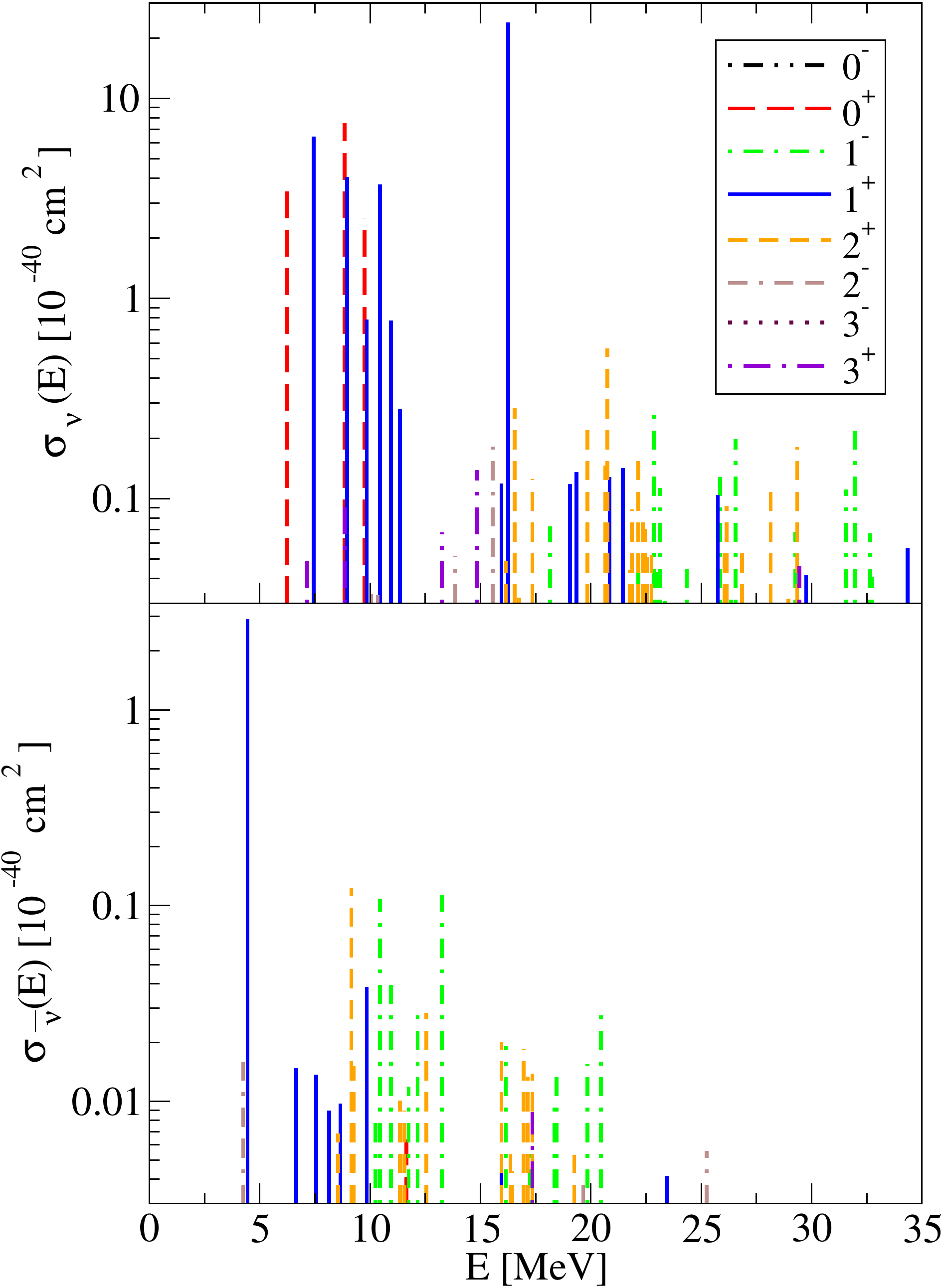}
}
\caption{{Exclusive cross sections for $\nu_{e}$ - $^{56}$Fe (upper panel) and 
$\bar{\nu}_{e}$ - $^{56}$Fe (lower panel) reactions as a function of excitation 
energy in initial nuclei, including $J=0^{\pm}-3^{\pm}$. }}
\label{fig1}
\end{figure}

For the purpose of this work, first we determine the cross sections for the 
charged current $\nu_{e}(\bar{\nu}_{e})$ - nucleus reactions for the following target 
nuclei: $^{12}$C, $^{16}$O, $^{56}$Fe and $^{208}$Pb. 
{The exclusive cross sections are described in the RNEDF 
framework~\cite{Paa.08,Paa.11}, 
by employing the density dependent effective interaction DD-ME2 \cite{Lal.05}.
Transition matrix elements for neutrino-induced reactions are calculated using
the general formalism from Refs.~\cite{Con.72, Wal.75}.
This method allows to determine in a consistent way the ground state properties of 
target nuclei and transitions induced by neutrinos.}
The exclusive cross sections are calculated as functions of excitation energy of initial nuclei, 
including all contributions from the initial ground state of even-even nucleus to the 
particular excited state, for all relevant multipolarities, $J<=5$, and both parities.
Figure~\ref{fig1} shows the cross sections of $\nu_{e}(\bar{\nu}_{e})$ - $^{56}$Fe reactions with
charge exchange, displayed as a function of the excitation energy in initial nucleus, 
for various multipoles, $J^{\pi}=0^{\pm}-3^{\pm}$.
The cross section of $\nu_e$ - $^{56}$Fe CC reactions is mainly determined by
$J^{\pi}=0^{+},1^{+}$ transitions, while the contribution of $J^{\pi}=1^{-},2^{-}$ states
is an order of magnitude smaller. Other multipolarities contribute only marginally. 
As shown in Fig.~\ref{fig1}, in the case of $\bar{\nu}_e$ - $^{56}$Fe reactions the
cross section is dominated by $J^{\pi}=1^{+}$ transitions, with major contribution from
the excited state at 4.5 MeV. In the limit of zero momentum
transfer, {the respective GT$^{\pm}$ transitions}
result in reasonable agreement with respective experimental data \cite{Paa.11, Niu.11}.
Energy threshold for CC reactions in $^{56}$Fe is relatively low for $\nu_e$ (4.57 MeV) 
and $\bar{\nu}_e$ (4.72 MeV).
The $\bar{\nu}_e$ - $^{56}$Fe cross sections are an order of magnitude smaller
than in the neutrino case (Fig~\ref{fig1}) due to the {effect of Pauli blocking of 
neutron quasiparticle states.}
%

\begin{figure}
\centerline{
\includegraphics[scale=0.35,angle=0]{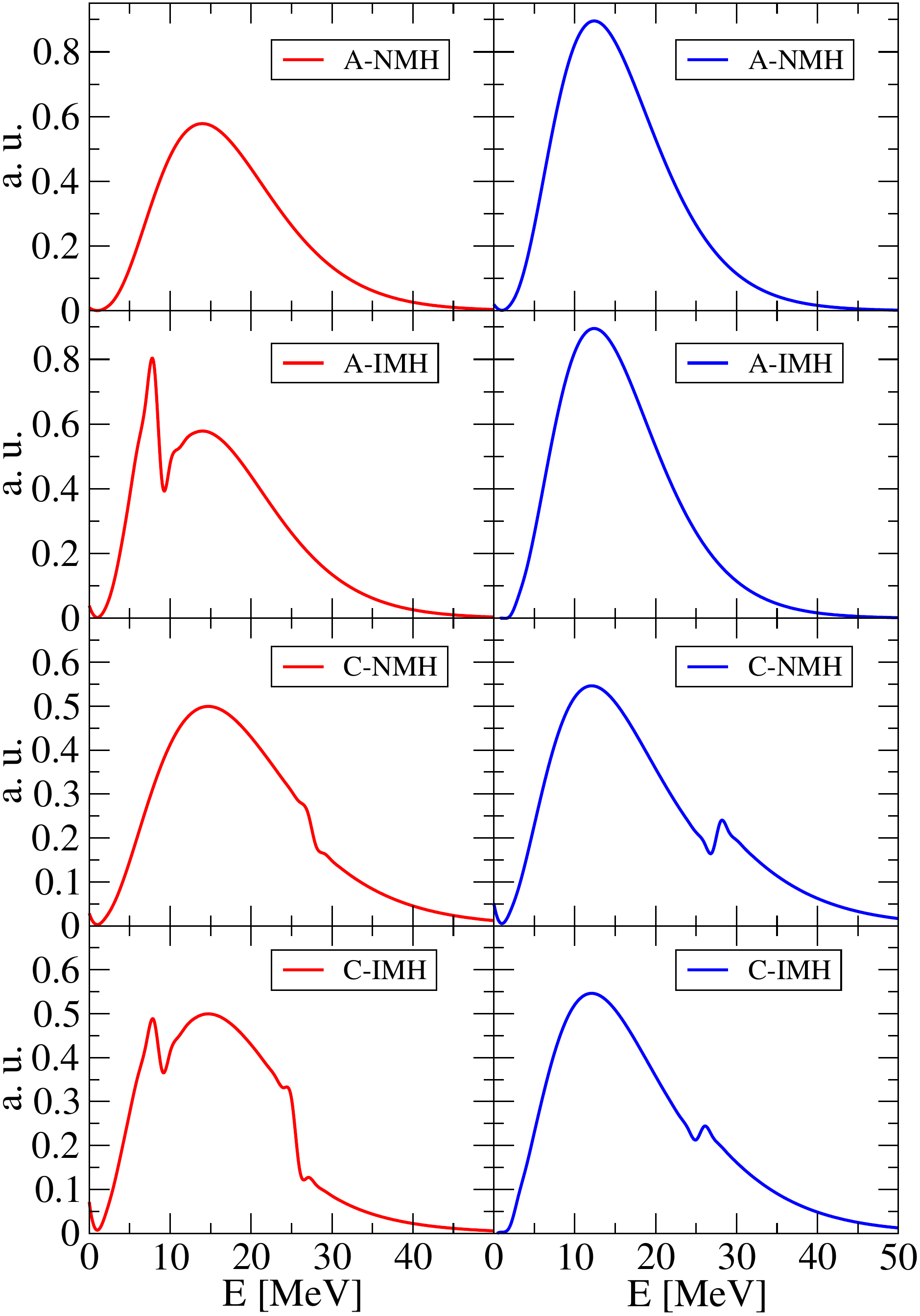}
}
\caption{ Incoming $\nu_e$ (left side) and $\bar{\nu}_e$ fluxes (right side) 
for the accretion (A) and cooling (C) phase of type II supernovae as a function
of neutrino energy for normal (NMH) and inverted neutrino mass hierarchy (IMH). }
\label{fig2}
\end{figure}

Initial $\nu_e$($\bar{\nu}_e$) spectra of the type II supernova can be described in the 
quasistationary approximation by the Fermi-Dirac or power law distribution \cite{Vaa.11, Mir.11, Das.10}.  {One should note} that initial neutrino fluxes are 
variable in time \cite{Jan.07}. The fits to the available data from SN1987A have shown that average neutrino energy is relatively low, i.e. $\sim$10 MeV \cite{Vaa.11}.
In the present work, the power law distribution is used to 
describe initial supernova neutrino spectra at neutrinospheres 
with fixed value of spectral parameter, i.e. 
4.0 in the case of accretion neutrino flux and 3.0 during the whole cooling phase~\cite{Hud.11}.
In numerical simulations of the neutrino accretion 
fluxes we fixed initial average energies of neutrinos to 
12, 15 and 18 MeV for $\nu_e$, $\bar{\nu}_e$, and non-electron 
species, respectively~\cite{Cho.10, Fis.10, Mir.11}. Certain deviations of previous values can 
be found in other supernova simulations~\cite{Hud.11}. In this work the 
best-fit values of neutrino oscillation parameters are used \cite{Fog.12}, except in 
region of collective oscillations where matter suppressed values of neutrino mixing are used instead \cite{Mir.11}.
For the initial accretion luminosities 
of  (anti)neutrino species we take 2.4 $\times 10^{52}$ ergs in {the case of $\nu_e$, 
2.0 $\times 10^{52}$ ergs for $\bar{\nu}_e$}, and $10^{52}$ ergs for
all other (anti)neutrino species. In the supernova cooling phase neutrino luminosities 
are almost an order of magnitude smaller, i.e. 
the initial values we used are 1.2 $\times 10^{51}$ ergs for $\nu_e$($\bar{\nu}_e$)
and 1.8 $\times 10^{51}$ ergs for non-electron species, and the initial average energies of non-electron species are 
assumed 2 MeV higher than in 
the accretion phase \cite{Cho.10}. The incoming (anti)neutrino fluxes are calculated including 
collective and MSW effects in the core-collapse star. 
Incoming $\nu_e$ and $\bar{\nu}_e$ fluxes of accretion and cooling supernova 
phase for normal (NMH) and inverted mass hierarchy (IMH) are shown in Fig.~\ref{fig2}. 

{The calculated $\nu_e$($\bar{\nu}_e$) - nucleus cross sections are used to
determine the respective flux averaged values  by employing $\nu_e$($\bar{\nu}_e$) distributions shown in Fig.~\ref{fig2}, and to evaluate the total number of 
detector events. 
In addition to $\nu_{e}(\bar{\nu}_{e})$ - nucleus cross sections, inelastic
$\bar{\nu}_{e}$ scattering on free protons is taken into account in mineral oil {(alkanes based on $-\text{CH}_{2}-$ group)}
and water ($\text{H}_2\text{O}$).  We employ the theory framework  based on heavy-barion
chiral perturbation theory  which also
includes radiative corrections ~\cite{Rah.12}.
It is assumed that the detector efficiency is perfect, and it is turned directly to the incoming neutrino flux, thus 
eliminating the Earth effects} on neutrino spectra. 
Four cases of target material are considered, mineral oil ($\text{CH}_{2}$), water (H$_{2}$O), $^{56}$Fe, and $^{208}$Pb.
As a test case for type II supernova {in our galaxy}, an imaginary star 12000 l.y. away from {the Earth} is 
considered. {We have calculated several neutrino induced decay channels, including $\gamma$, one-neutron (1n) and 
two-neutron (2n) emission, based on the limits given by the separation energies $S_{1n}$, $S_{2n}$, $S_{1p}$, 
calculatated from the implementation of 
the RNEDF in the relativistic Hartree-Bogoliubov model~\cite{Lal.05}}. 
{Table~\ref{table1} shows the number 
of $\nu_e$($\bar{\nu}_{e}$) - nucleus events}
({only dominant decay channels are shown}) in 1 kt of target material based on
$^{56}$Fe and $^{208}$Pb, and the number of emitted neutrons in 
case of $\bar{\nu}_{e}-p$ reaction in mineral oil ($\text{CH}_{2}$) and water ($\text{H}_{2}\text{O}$).
{Calculations include 
incoming $\nu_e$($\bar{\nu}_{e}$) fluxes of the accretion and cooling supernova phase, both for 
normal and inverted neutrino mass hierarchies.} For the 
time duration of accretion and cooling supernova phase is taken 0.2 s and 10 s, 
respectively \cite{Fis.10}.
%
\begin{table}
\caption{{Detector response for $\nu_{e}$($\bar{\nu}_{e}$) - induced reactions in mineral oil (CH$_{2}$), water (H$_{2}$O), $^{56}$Fe and $^{208}$Pb,
 for the incoming (anti)neutrino fluxes of the accretion (A) and cooling supernova phase (C), both for
normal (NMH) and inverted (IMH) neutrino mass hierarchies.
Only dominant emission channels are shown, including $\gamma$ rays, 1n and 2n emissions.}}
\begin{center}
\item[]\begin{tabular}{@{}*{5}{l}}
\hline
$ $ & \multicolumn{2}{c}{NMH} & \multicolumn{2}{c}{IMH} \cr
$ $&$N(A)$&$N(C)$&$N(A)$&$N(C)$\cr
\hline
$p(\bar{\nu}_{e},e^-)n$ in CH$_{2}$\cr
1n&160&900&160&815\cr
\hline
$p(\bar{\nu}_{e},e^-)n$ in H$_{2}$O\cr
1n&125&700&125&635\cr
\hline
$^{56}$Fe$(\nu_{e},e^-)^{56}$Co& & & &\cr
$\gamma$&19&115&19&80\cr
1n&17&137&17&80\cr
\hline
$^{56}$Fe$(\bar{\nu}_{e},e^+)^{56}$Mn& & & &\cr
$\gamma$&2&19&2&16\cr
\hline
$^{208}$Pb$(\nu_{e},e^-)^{208}$Bi& & & &\cr
$\gamma$&12&70&12&51\cr
1n&84&533&84&350\cr
2n&11&120&11&59\cr
\hline
\end{tabular}
\end{center}
\label{table1}
\end{table}
The results
indicate a higher number of the events related to $\nu_{e}$($\bar{\nu}_{e}$)
in the case of NMH. 
Detector response of accretion $\nu_{e}$($\bar{\nu}_{e}$) flux is characterized by the degeneracy  
in emission rates of $\gamma$ rays, protons and neutrons for the entire set of target material 
and the best-fit values of neutrino oscillation parameters \cite{Fog.12}, 
i.e. it is invariant on the type of mass hierarchy. Similar result is obtained in 
simulations of supernova neutrino fluxes in case of $^{208}$Pb and 
large $\theta_{13}$ mixing in \cite{Vaa.11}. 

The most pronounced response to CC interactions with neutrinos 
is obtained for $^{208}$Pb. It has large number of protons $(Z=82)$, so the 
Coulomb effects enchance the phase 
space for emitted electrons \cite{Ful.99}, large neutron excess $(N-Z=44)$ 
and relatively low treshold (2.88 MeV). The expected 
number of $\gamma$ and proton events due to deexcitation of 
$^{208}$Bi is similar for both neutrino mass hierarchies.
Within the HALO detector \cite{Dub.08}, supernova neutrinos
will be observed through neutron emissions from Pb target
using $^{3}$He neutron detectors.
For the supernova cooling phase, we obtain the flux averaged cross section 
related with 1n-emission from $^{208}$Bi: $5.6\times10^{-40}$ cm$^2$ in the case of NMH, 
and $3.8\times10^{-40}$ cm$^2$ in the case of IMH. However, significant contribution
of 2n events due to relatively high 2n cross section rates is also obtained,
i.e., $1.2\times10^{-40}$ cm$^2$ in the case of NMH and $0.6\times10^{-40}$ cm$^2$ 
in the case of IMH. The resulting total number of emitted neutrons amounts
$\approx$ 880 in the case of NMH, and $\approx$ 575 in the case of IMH, 
with the ratio of emitted neutrons $N_{n}^{NMH}/N_{n}^{IMH}\approx1.5$. 
{The sensitivity of 2n-emissions on the type of mass hierarchy is also 
obtained, i.e., $N_{2n}^{NMH}/N_{2n}^{IMH}\approx2.0$.}
The cross sections for $\bar{\nu}_{e} - ^{208}$Pb reactions are strongly 
suppressed (3 orders of magnitude) due to Pauli blocking of the neutron
single-particle states. 

The case of $^{56}$Fe {target} is characterized 
by somewhat weaker response to CC reactions with neutrinos (Tab.~\ref{table1}), with
almost half predicted events coming from $^{56}$Co $\gamma$ decay, while the other half 
mainly comes from 1n emission. The 1p and 2n emissions are 
severely reduced. However, due to smaller neutron excess ($N-Z=4$), detector based
on $^{56}$Fe is also sensitive to antineutrino CC reactions with respective cross 
sections $\sim10^{-42}$cm$^2$, an order of magnitude smaller than for $\nu_e - ^{56}$Fe reaction.
The predicted total cross section of neutrino induced reaction in $^{56}$Fe is 
$\sim$10 times smaller than for $^{208}$Pb. The total number of 
neutrons emitted from the detector (for $\nu_{e}$ and $\bar{\nu}_{e}$) based on $^{56}$Fe is $\approx$ 165 in the case of NMH and 
$\approx$ 110 in case of IMH, i.e., it is $\approx1.5$ times larger in the case of NMH.

In the case of mineral oil (water) of density $0.85$ g/cm$^3$ (1.0 g/cm$^3$ at 4 $^\circ$C), 
where the target nuclei are $^{12}$C ($^{16}$O), the predicted cross sections are of the order $\sim10^{-42}$ cm$^{2}$, and due to $N=Z=6$ $(N=Z=8)$ the difference in total number of primary events between neutrino and antineutrino CC reactions is smaller than for $N>Z$ nuclei.
{Due to relatively high energy thresholds ($\gtrsim11 \text{ MeV}$) for both types of reactions in $^{12}$C ($^{16}$O), at 
$\nu_{e}$ ($\bar{\nu}_{e}$) energies $\lesssim20 \text{ MeV}$ the responses are rather low.} 
{Actually, in the case of mineral oil (water) $\bar{\nu}_{e}$ induce reactions mainly with 
free protons, resulting in 1060 (825) events in the case of NMH and 975 (760) 
events in the case of IMH, respectively (Tab.~\ref{table1}).}
{Dominance of $\bar{\nu}_e-p$
reactions in mineral oil and water} due to low energy 
threshold (1.8 MeV) and $\sigma\sim10^{-43}(E_{\nu}/MeV)^{2}$ {ensures an efficient coverage} 
of $\bar{\nu}_{e}$ spectra. 

{The results of the present analysis show that neither water 
or mineral oil can provide an evidence to determine neutrino mass hierarchy 
due to rather small difference of $p(\bar{\nu}_e,e^{+})n$ events for NMH and IMH fluxes (Tab.~\ref{table1}).}
Therefore, we implement {a hybrid approach, based on}
combination of detectors with different target materials 
to cover complementary parts of spectra in order to calculate relevant ratios of the events 
that are sensitive on the neutrino 
mass hierarchy. {By focusing only on the} cooling phase fluxes, the 
respective ratios {of the neutron emissions for NMH and IMH} in the case of combination of two detectors, 
1 kt of water (free protons only) and 1 kt of $^{208}\text{Pb}$ are: 
$N_{n}^{NMH}(\text{H}_{2}\text{O})/N_{n}^{NMH}(\text{Pb})=(0.91\pm0.05)$ 
and $N_{n}^{IMH}(\text{H}_{2}\text{O})/N_{n}^{IMH}(\text{Pb})=(1.36\pm0.08)$. The statistical
error of {predicted} events is estimated as $\sqrt{N}$.
{In addition}, one can also use time-difference of accretion (A)
and cooling (C) phase signal, and accretion degeneracy. Greater difference of ratios of emitted neutrons, i.e. 
$N_{C}^{NMH}(\text{Pb})/N_{A}^{NMH}(\text{Pb})=(7.3\pm0.8)$ 
and $N_{C}^{IMH}(\text{Pb})/N_{A}^{IMH}(\text{Pb})=(4.4\pm0.5)$, once again confirms $^{208}\text{Pb}$ 
as preferred target nucleus {to cover the neutrino part of the spectra}.

In conclusion, {we have presented a hybrid approach} to determine neutrino mass
hierarchy from {simultaneous} measurements of supernova $\nu_{e}$ and 
$\bar{\nu}_{e}$ time difference in the events in detectors based on different types of target material.
 {Through the use of supernova $\nu_{e}$($\bar{\nu}_{e}$) fluxes that include
 collective and MSW effects for the 
accretion and cooling phase, responses in mineral oil, water, $^{56}$Fe and
$^{208}$Pb have been analyzed for both possible neutrino mass hierarchies.}
{It is shown that the hybrid method, that combines antineutrino induced events
in water or mineral oil, with neutrino induced emissions in heavier target nuclei
such as $^{208}$Pb, provides a useful diagnostic tool to constrain the neutrino
mass hierarchy. The number of 
emitted neutrons for the case of NMH is $\approx1.5$ larger than for IMH, both
for $^{208}$Pb and $^{56}$Fe target nuclei. Since it is rather difficult to compare
the absolute values of calculated particle emissions with the detector events, we 
have shown that the relative ratio of  antineutrino and neutrino induced events 
in different detectors represent a quantity that could discriminate between 
different neutrino mass hierarchies.}
{It is important to note that  heavy nuclei are almost completely inert for CC reactions 
with antineutrinos due to strong Pauli blocking of neutron single particle states. Pure neutrino signal and 
large cross section establish $^{208}$Pb as probably the most 
important nuclear target for neutrino detection and reconstruction of the
neutrino part of spectra.}
{In case of the occurrence of the next galactic supernova, and taking
into account current neutrino detector developments (e.g., HALO ($^{208}$Pb)~\cite{Dub.08}, Super-Kamiokande (water)~\cite{Kam.14}, etc.)
the hybrid method presented in this work would play an important role in constraining
the neutrino mass hierarchy.}
%

%

\end{document}